\documentclass[12pt,a4paper]{article}
\usepackage{amssymb}
\usepackage{graphicx}
\usepackage{graphics}
\usepackage{epsfig}
\usepackage{amsmath}
\usepackage{url}
\usepackage{multicol}
\usepackage{float}
\usepackage{empheq}
\begin{document}

\begin{center}
{\Large \bf Some properties of U(1) gauged Q-balls}

\vspace{4mm}

I.\,E.\;Gulamov$^{a,b}$, E.\,Ya.\;Nugaev$^c$, A.\,G.\;Panin$^c$,
M.\,N.\;Smolyakov$^b$\\
\vspace{0.5cm} $^a${\small{\em Physics Department, Lomonosov
Moscow State University,}}\\
{\small{\em 119991, Moscow, Russia}}\\
$^b${\small{\em Skobeltsyn Institute of Nuclear Physics, Lomonosov
Moscow State University,}}\\
{\small{\em 119991, Moscow, Russia}}\\
$^c${\small{\em Institute for Nuclear Research of the Russian
Academy of Sciences,}}\\
{\small{\em 60th October Anniversary prospect 7a, 117312, Moscow,
Russia}}
\end{center}

\begin{abstract}
In this paper we examine the properties of $U(1)$ gauged Q-balls in two models with different scalar field potentials. The obtained results demonstrate that in the general case $U(1)$ gauged Q-balls possess properties, which differ considerably from those of Q-balls in the nongauged case with the same forms of the scalar field potential. In particular, it is shown that in some cases the charge of $U(1)$ gauged Q-ball can be bounded from above, whereas it is not so for the corresponding nongauged Q-ball. Our conclusions are supported both by analytical considerations and numerical calculations.
\end{abstract}

\section{Introduction}
A simplest generalization of nontopological solitons, initially proposed in \cite{Rosen0} and known as Q-balls \cite{Coleman:1985ki}, from the global $U(1)$ symmetry to the gauge $U(1)$ symmetry was proposed and analyzed in the pioneering paper \cite{Rosen}. Later this subject was examined in the well-known paper \cite{Lee:1988ag}, in which gauged Q-balls (for simplicity, from here on, we call $U(1)$ gauged Q-balls ``gauged Q-balls'') were examined analytically and numerically. One can also recall papers \cite{BF,BF1,Gulamov:2013cra}, where gauged Q-balls were examined mainly from a theoretical point of view, as well as papers \cite{Lee:1991bn,Arodz:2008nm,Dzhunushaliev:2012zb,Tamaki:2014oha,Brihaye:2014gua}, where solutions for gauged Q-ball were obtained numerically.

It is clear that if the backreaction of the gauge field on the scalar field is small, the characteristics of gauged Q-balls (charge, energy, etc.) do not differ considerably from those of ordinary nongauged Q-balls (see paper \cite{Gulamov:2013cra}, in which this issue was examined in detail). Meanwhile, the most interesting cases are those in which backreaction of the gauge field cannot be neglected. In the general case, the latter realizes not only if the corresponding coupling constant is large, but even when the coupling constant is rather small, but the other parameters of the solution are such that contribution of the gauge field is considerable. In the present paper we examine, both analytically and numerically, such differences between nongauged and gauged cases. In particular, we show that in some cases gauged Q-balls may exist for such values of the parameters, for which nongauged Q-balls do not exist at all. And vice versa, it is possible that there are no gauged Q-balls for the values of the parameters for which nongauged Q-balls exist. The obtained results demonstrate that gauged Q-balls possess properties, which can be completely different from those of Q-balls in the nongauged case.

The paper is organized as follows. In Section~2 we present the
general setup and introduce the notations that will be used throughout
the paper. In Section~3 we present some analytical results related to
gauged Q-balls. In Section~4 we examine numerically two models with different scalar field potentials. The obtained results are briefly discussed in the last section.

\section{Setup}
We consider the action, describing the simplest $U(1)$ gauge
invariant scalar field theory in four-dimensional space-time, in the form
\begin{equation}\label{action}
S=\int
d^4x\left((\partial^{\mu}\phi^{*}-ieA^{\mu}\phi^{*})(\partial_{\mu}\phi+ieA_{\mu}\phi)-V(\phi^{*}\phi)-\frac{1}{4}F_{\mu\nu}F^{\mu\nu}\right)
\end{equation}
and take the standard spherically symmetric ansatz for the scalar and gauge fields \cite{Rosen,Lee:1988ag}:
\begin{eqnarray}\label{ans1}
\phi(t,\vec x)&=&\textrm{e}^{i\omega t}f(r),\qquad
f(r)|_{r\to\infty}\to 0, \qquad \frac{df(r)}{dr}\biggl|_{r=0}=0,\\
\label{ans2}A_{0}(t,\vec x)&=&A_{0}(r),\qquad\,\,
A_{0}(r)|_{r\to\infty}\to 0,\qquad
\frac{dA_{0}(r)}{dr}\biggl|_{r=0}=0,\\ \label{ans3} A_{i}(t,\vec
x)&\equiv &0,
\end{eqnarray}
where $r=\sqrt{\vec x^{2}}$ and $f(r)$, $A_{0}(r)$ are real
functions. Below we will consider solutions such that the function $f(r)$ has no nodes. Without loss of generality we take
$f(0)>0$.

It is obvious that, according to (\ref{ans1})--(\ref{ans3}), we can use the
effective action
\begin{equation}\label{effaction}
S_{\textrm{eff}}=4\pi\int\limits_{0}^{\infty}
r^{2}dr\left((\omega+eA_{0})^{2}f^2-\partial_{r}f\partial_{r}f-V(f)+\frac{1}{2}\partial_{r}A_{0}\partial_{r}A_{0}\right),
\end{equation}
where $V(f)=V(\phi^{*}\phi)$. For the scalar field potential, the
conditions
\begin{equation}
V(0)=0,\qquad \frac{dV}{df}\biggl|_{f=0}=0
\end{equation}
are supposed to fulfill in order to ensure the existence of the vacuum solution $f(r)\equiv 0$, $A_{0}(r)\equiv 0$. The equations of motion, following from effective action
(\ref{effaction}), take the form
\begin{eqnarray}\label{eqg1}
2e(\omega+eA_{0})f^2=\frac{1}{r}\frac{d^{2}}{dr^{2}}(rA_{0}),\\
\label{eqg2} (\omega+eA_{0})^2f+\frac{1}{r}\frac{d^{2}}{dr^{2}}(rf)-\frac{1}{2}\frac{dV}{df}=0.
\end{eqnarray}
For the numerical analysis, it is more convenient to use the combination $a(r)=\omega+eA_{0}(r)$ instead of the field $A_{0}(r)$. With this notation, equations (\ref{eqg1}), (\ref{eqg2}) can be rewritten as
\begin{eqnarray}\label{eqg1a}
2e^2af^2=\frac{1}{r}\frac{d^{2}}{dr^{2}}(ra),\\
\label{eqg2a} a^2f+\frac{1}{r}\frac{d^{2}}{dr^{2}}(rf)-\frac{1}{2}\frac{dV}{df}=0,
\end{eqnarray}
where $\frac{da(r)}{dr}\bigl|_{r=0}=0$. The value of the frequency $\omega$ is now defined as $\omega=\lim\limits_{r\to\infty}a(r)$.

The charge of a gauged Q-ball can be defined as\footnote{The physical charge is defined by $Q_{phys}=eQ$, but below we will use the charge $Q$ defined by (\ref{chargedef}), which simplifies comparison with the nongauged case.}
\begin{equation}\label{chargedef}
Q=8\pi\int\limits_{0}^{\infty}(\omega+eA_{0})f^2r^2dr=8\pi\int\limits_{0}^{\infty} af^2 r^2dr.
\end{equation}
According to \cite{Rosen,Lee:1988ag}, the sign of $a=\omega+eA_{0}$ always coincides with the sign of
$\omega$, whereas $A_{0}\equiv 0$ for $\omega=0$. Thus, without loss of generality we can consider $\omega\ge 0$,
which leads to $Q\ge 0$.
The energy of a gauged Q-ball at rest is defined by
\begin{equation}\label{energydef}
E=4\pi\int\limits_{0}^{\infty}
\left(a^2f^2+\partial_{r}f\partial_{r}f+V(f)+\frac{1}{2e^2}\partial_{r}a\partial_{r}a\right)r^2dr.
\end{equation}

It is well known that for ordinary (nongauged) Q-balls the relation $\frac{dE}{dQ}=\omega$ holds. In \cite{Gulamov:2013cra} it was shown that the same relation also holds for $U(1)$ gauged Q-balls. We will use it for an extra check of our numerical results.

\section{Analytical considerations}
To begin with, let us discuss the allowed values of the frequency $\omega$. Suppose that our scalar field potential is such that the relation
\begin{equation}
\frac{1}{2f}\frac{dV}{df}\biggl|_{f=0}=M^{2}
\end{equation}
holds. In the most cases nongauged (i.e., with global $U(1)$ symmetry) Q-balls exist only for $\omega<M$, whereas $Q\to\infty$ for $\omega\to M$. So, it is universally accepted that for gauged Q-balls the values of the frequency $\omega$ are also bounded from above as $\omega<M$ (recall that we take $\omega\ge 0$). Indeed, it was shown in \cite{Rosen} using the perturbation method in the effective coupling constant that the total energy of gauged Q-ball diverges for $\omega=M$ even in the special case in which the corresponding nongauged Q-ball exists and has finite charge and energy\footnote{Note that the correction to the background nongauged solution in \cite{Rosen} grows with $r$, which indicates the breakdown of the linear approximation at some $r$ (moreover, as it was shown in \cite{Gulamov:2013cra}, such a breakdown of the linear approximation for the correction is inherent to models of gauged Q-balls). Thus, the divergence of total energy in the linear approximation cannot be used as an indication of the absence of a solution to the full set of nonlinear equations.}. In \cite{Lee:1988ag} it was stated that the condition $\omega<M$ is required to have localized solutions without oscillations for the scalar field. However, our numerical analysis shows that gauged Q-balls with finite charge and energy for $\omega=M$ may exist even if the corresponding solution in the nongauged case does not exist at all. Below we will present some analytical considerations which support this statement.

First, we consider the usual case $\omega<M$ and suppose that there exists a gauged Q-ball solution with finite charge and energy such that $A_{0}(r)\to-\frac{eQ}{4\pi r}$ for large $r$. For such large $r$ the equation for the scalar field can be rewritten as
\begin{equation}\label{lineqf}
(\omega^2-M^2)f-\frac{2\,\omega e^{2}Q}{4\pi r}f+\frac{1}{r}\frac{d^{2}}{dr^{2}}(rf)\approx 0.
\end{equation}
The solution to this equation, tending to zero as $r\to\infty$, can be easily obtained and takes the form
\begin{equation}\label{frinfty}
f(r)=C\textrm{e}^{-\sqrt{M^2-\omega^2}\,r}U\left(1+\frac{\omega e^2Q}{4\pi\sqrt{M^2-\omega^2}},2,2\sqrt{M^2-\omega^2}\,r\right),
\end{equation}
where $C$ is a constant and $U(b,c,z)$ is the confluent hypergeometric function of the second kind. It is not difficult to show (see Appendix~A) that for $\sqrt{M^2-\omega^2}\,r\gg 1$ we get for (\ref{frinfty})
\begin{equation}\label{frinfty1}
f(r)\sim \frac{\textrm{e}^{-\sqrt{M^2-\omega^2}\,r}}{r^{1+\frac{\omega e^2Q}{4\pi\sqrt{M^{2}-\omega^2}}}}.
\end{equation}
This formula resembles the naively expected result $f(r)\sim\frac{\textrm{e}^{-\sqrt{M^2-\omega^2}\,r}}{r}$, the difference is caused by taking into account the electromagnetic potential $A_{0}(r)\to-\frac{eQ}{4\pi r}$ for large $r$. But formula (\ref{frinfty1}) has a singular behavior in the limit $\omega\to M$, which clearly indicates that the case $\omega=M$ should be considered separately.

Equation (\ref{lineqf}) for $\omega=M$ also has a solution, ensuring the finiteness of charge and energy. Indeed, let us suppose that solution to the full set of nonlinear equations exists and has the charge $Q$. Then, far away from the center of the Q-ball, we can write for the scalar field
\begin{equation}\label{lineqf2}
-\frac{2Me^{2}Q}{4\pi r}f+\frac{1}{r}\frac{d^{2}}{dr^{2}}(rf)\approx 0.
\end{equation}
The solution to this equation, tending to zero as $r\to\infty$, takes the form
\begin{equation}\label{frinfty2}
f(r)=C\frac{K_{1}\left(\sqrt{\frac{2Me^{2}Q}{\pi}\,r}\right)}{\sqrt{r}},
\end{equation}
where $C$ is a constant and $K_{1}(b,z)$ is the modified Bessel function of the second kind. For large $r$ this solution behaves as
\begin{equation}\label{frinfty1a}
f(r)\sim \frac{\textrm{e}^{-\sqrt{\frac{2Me^{2}Q}{\pi}\,r}}}{r^{\frac{3}{4}}}.
\end{equation}
On sees that due to the long-range action of the electromagnetic potential $A_{0}(r)$, the behavior of the scalar field at $r\to\infty$ differs considerably from the nongauged case, in which one expects $f(r)\sim\frac{1}{r}$ for $\omega=M$. Moreover, one may naively expect that the repulsive nature of the electrostatic interaction would prevent from forming a gauged Q-ball for such a value of $\omega$ (recall that usually the corresponding nongauged Q-ball has an infinite charge for $\omega=M$). The argumentation presented above shows that it is not so. This statement is also confirmed by the numerical results, which will be presented in the next section.

It should be noted that it is possible to show analytically that in the limit $\omega\to M$ solution (\ref{frinfty}) transforms into solution (\ref{frinfty2}), see Appendix~B.

A few words about the case $\omega>M$. The form of equation (\ref{lineqf}) suggests that the corresponding solutions for the scalar field are oscillatory for $r\to\infty$, leading to infinite charge and energy. More precisely, the leading term of a solution to equation (\ref{lineqf}) takes the form
\begin{eqnarray}\nonumber
C_{1}\frac{1}{r}\cos\left(\sqrt{\omega^2-M^2}\,r-\frac{\omega e^2Q}{4\pi\sqrt{\omega^2-M^2}}\ln\left(\sqrt{\omega^2-M^2}\,r\right)\right)\\ \nonumber+C_{2}\frac{1}{r}\sin\left(\sqrt{\omega^2-M^2}\,r-\frac{\omega e^2Q}{4\pi\sqrt{\omega^2-M^2}}\ln\left(\sqrt{\omega^2-M^2}\,r\right)\right).
\end{eqnarray}
The latter formula supports the assumption that there are no gauged Q-balls for $\omega>M$, which was also confirmed by the numerical analysis.

One can think that it looks rather strange that there exist solutions with finite charge for $\omega\le M$, whereas no solutions with finite charge are expected for $\omega>M$. However, a similar situation can be observed in the nongauged case. Indeed, let us take a scalar field potential of the form
\begin{equation}
V(f)=M^2f^2-\lambda |f|^N.
\end{equation}
It was noted in \cite{Multamaki:1999an} that $Q\to\textrm{const}\ne 0$ in the limit $\omega\to M$ for $N=\frac{10}{3}$ (one can easily check numerically that the corresponding solutions for the scalar field indeed exist). More precisely, $Q\sim\omega$ for $\omega<M$. Meanwhile, it is possible to show analytically that there are no Q-ball solutions for $\omega\ge M$ (to show it in a simple way one can use the scale transformation technique, proposed in \cite{Derrick}, supplemented by transformations of the fields themselves; see, for example, \cite{Smolyakov}, where such a method was applied to the more complicated case of gauged Q-balls). Thus, this example indicates that the situation with the existence of a finite charge for $\omega=M$ in the gauged case is not so unique.

\section{Explicit examples of $U(1)$ gauged Q-balls}
The numerical solutions for gauged Q-balls, which will be presented
below, were obtained in two steps. On the first step the shooting
method, which solves the boundary value problem by reducing it to
the solution of the initial value problem, was used. According to
this method, one should adjust the initial data for the system of
equations at one of the boundaries (at the origin of gauged Q-ball
in our case) in such a way that the solution satisfies required
conditions at the second boundary (for large $r$ in our case).
The shooting method is very simple and it is easy to implement it,
but it fails to find gauged Q-ball solutions for large $r$, where
the scalar field falls off to zero exponentially, see
relations~\eqref{frinfty1},~\eqref{frinfty1a}. To find a solution
for such large values of the coordinate $r$ using the shooting method,
one has to fine tune the initial data with very high accuracy, which
may even exceed the truncation error of double precision
floating-point numbers. Moreover, by taking the box of a small size one can mistake spurious solutions (such as oscillating solutions for $\omega>M$) for
correct monotonic solutions of the boundary value problem. So, these problems make the shooting method
not fully applicable for our task. In order to overcome them,
the solutions obtained on the first step (i.e., for the values of
$r$ which are not very large) were supplemented by analytical
solutions defined by formulas~\eqref{frinfty1},~\eqref{frinfty1a}. On the second
step, the resulted ``combined solutions'' were used as the
initial approximation of the solutions to the boundary
value problem for discretized version of the system of
equations~\eqref{eqg1a},~\eqref{eqg2a}. Then these solutions were
improved iteratively by the Gauss-Seidel red-black relaxations,
accelerated with the help of multigrid technique.
The additional cross-checks of the final results were performed
using the known theoretical relations for gauged Q-balls, such as,
for example, $\frac{dE}{dQ}=\frac{dE/d\omega}{dQ/d\omega}=\omega$~\cite{Gulamov:2013cra}.

\subsection{Model with $\phi^4$ scalar field potential}
At first, we consider the potential of form
\begin{equation}
V(f)=M^{2}f^{2}-\lambda f^{4},
\end{equation}
where $\lambda>0$. At the very beginning it is convenient to make the following redefinition of the coordinate $r$ and the fields:
\begin{equation}
R=Mr,\qquad G(R)=\frac{1}{M}\,a(r),\qquad F(R)=\frac{\sqrt{\lambda}}{M}f(r).
\end{equation}
In these notations, the system of equations (\ref{eqg1a}), (\ref{eqg2a}) takes the form
\begin{eqnarray}\label{eqg1a4}
2\alpha_{1}GF^2=\frac{1}{R}(RG)'',\\
\label{eqg2a4} G^2F+\frac{1}{R}(RF)''-F+2F^{3}=0,
\end{eqnarray}
where $'=\frac{d}{dR}$, $\alpha_{1}=\frac{e^2}{\lambda}$. We see that the only effective parameter in this system of equations is $\alpha_{1}$. Since $\frac{1}{2f}\frac{dV}{df}\bigl|_{f=0}=M^{2}$, we will be looking for solutions such that $G(\infty)\le 1$, which corresponds to $\omega\le M$.

\begin{figure}[!h]
\centering
\includegraphics[width=0.9\linewidth]{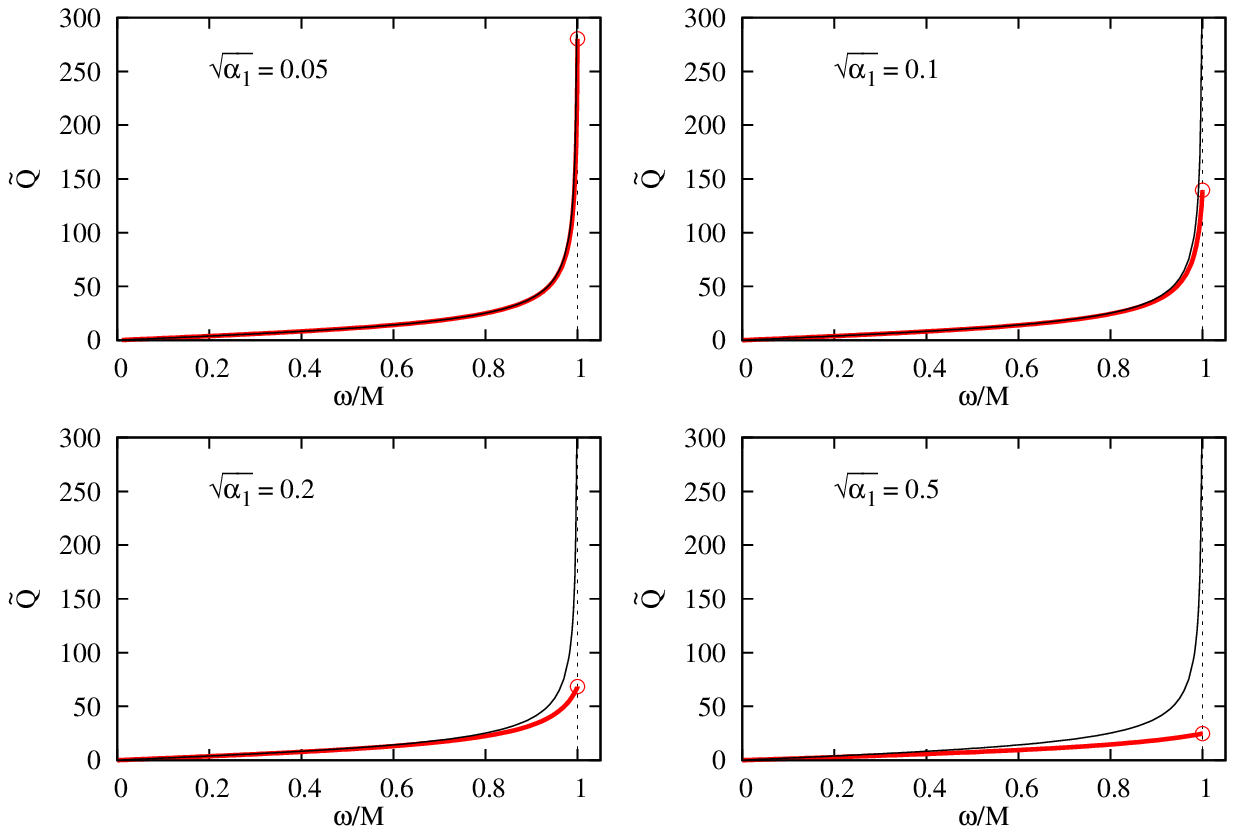}
\caption{$Q(\omega)$ for different values of the parameter $\alpha_{1}$ (thick lines). The thin lines stand for the nongauged case. The circles on the plots mark the points with $\frac{\omega}{M}=1$.}\label{fig1}
\end{figure}

The charge of the Q-ball takes the form
\begin{equation}\label{chargephi4}
Q=\frac{1}{\lambda}\,8\pi\int\limits_{0}^{\infty} GF^2R^2dR=\frac{1}{\lambda}\,\tilde Q,
\end{equation}
whereas the energy is
\begin{equation}\label{energyphi4}
E=\frac{M}{\lambda}\,4\pi\int\limits_{0}^{\infty}
\left(G^2F^2+\partial_{R}F\partial_{R}F+F^{2}-F^{4}+\frac{1}{2\alpha_{1}}\partial_{R}G\partial_{R}G\right)R^2dR=\frac{M}{\lambda}\,\tilde E.
\end{equation}

In Fig.~\ref{fig1} one can see several examples of $Q(\omega)$ diagrams (expressed in the dimensional variables $\tilde Q$ and $\frac{\omega}{M}$). Since in the nongauged case there exists a solution for $\omega=0$ with the zero charge \cite{AD}, the gauged solution for $\omega=0$ simply coincides with it \cite{Rosen}, which explains why the curves start from the point $\omega=0$, $\tilde Q=0$. Meanwhile, in the nongauged case the Q-ball charge tends to infinity while $\frac{\omega}{M}\to 1$ \cite{AD}. As it was demonstrated in Section~3, it is not so for the gauged case, in which one may expect the existence of a Q-ball with finite charge and energy.
\begin{figure}[!h]
\centering
\includegraphics[width=1\linewidth]{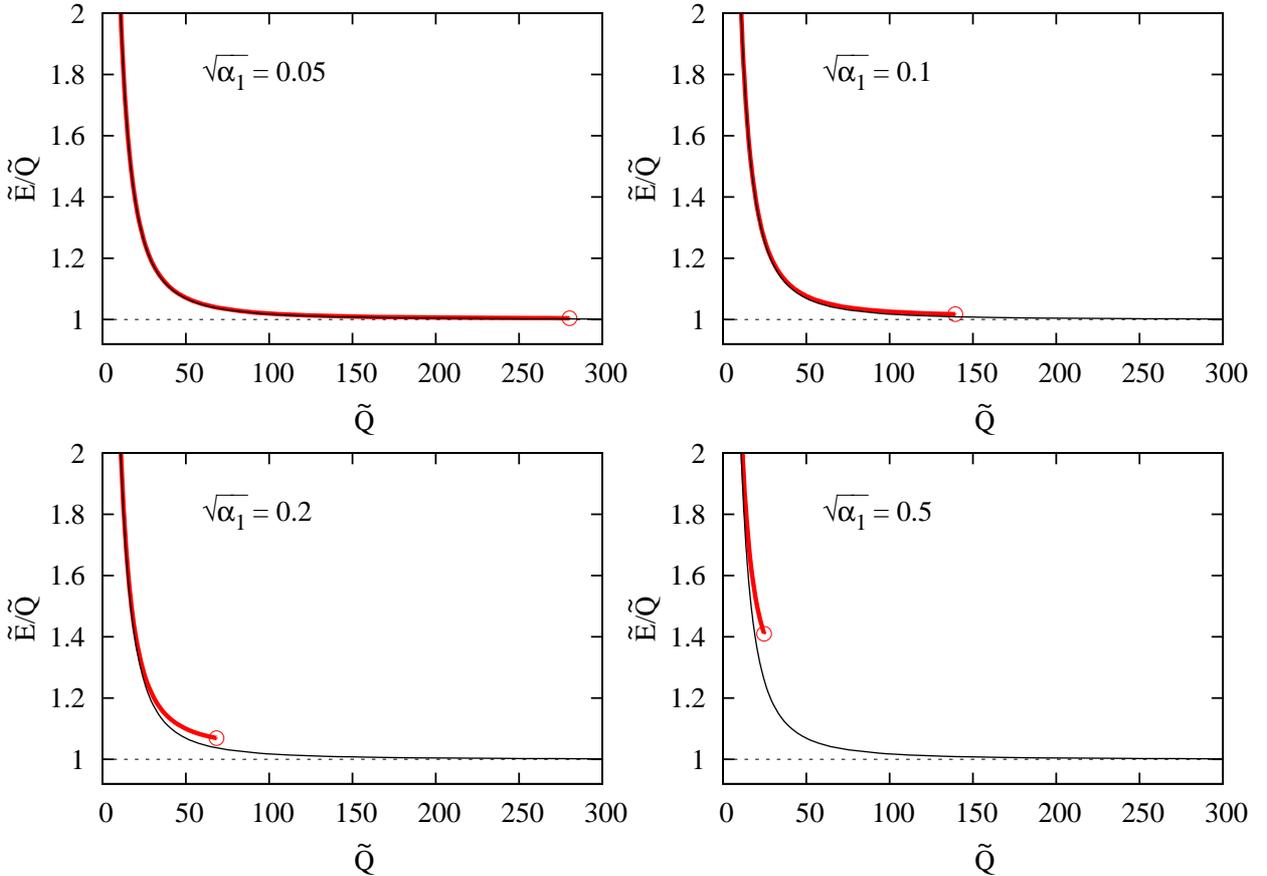}
\caption{$\frac{E}{Q}(Q)$ for different values of the parameter $\alpha_{1}$ (thick lines). The thin lines stand for the nongauged case. The circles on the plots mark the points with $\frac{\omega}{M}=1$.}\label{fig2}
\end{figure}
\begin{figure}[!ht]
\centering
\begin{minipage}[t]{0.49\textwidth}
\centering
\includegraphics[width=1\linewidth]{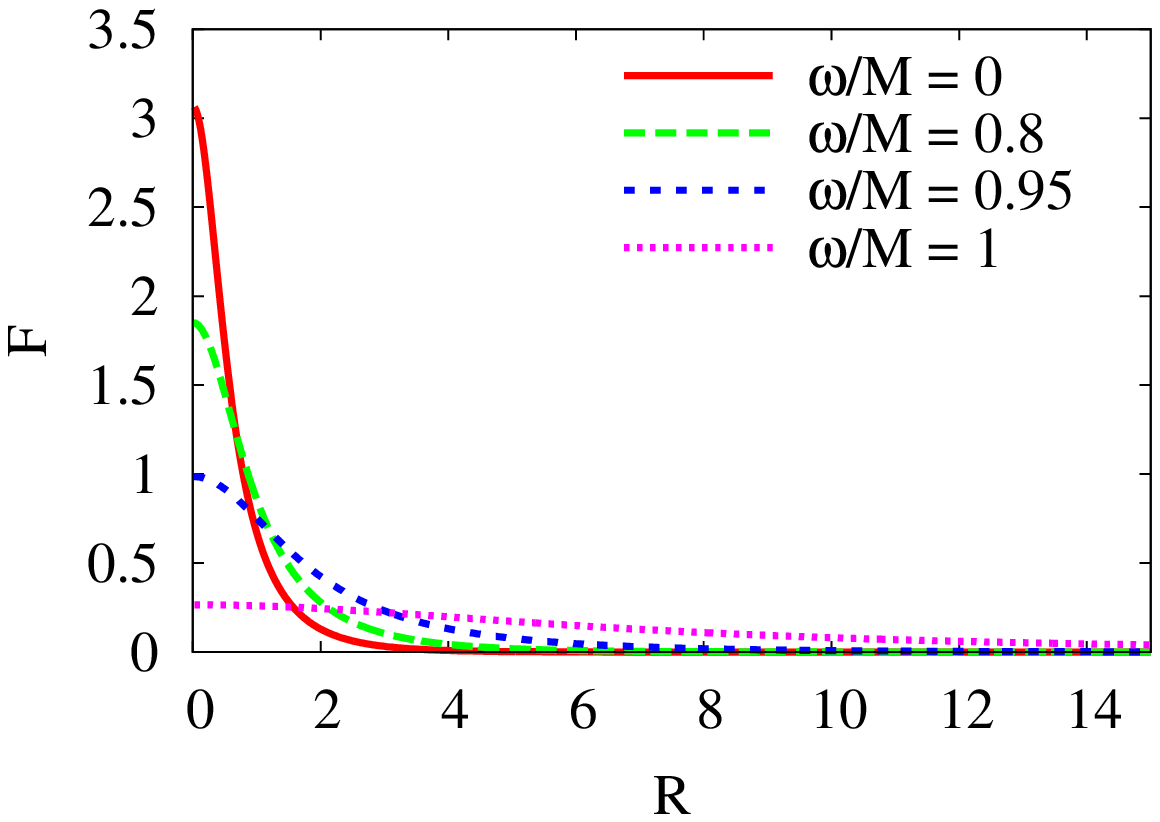}
\caption{Profiles of the scalar field for different values of $\frac{\omega}{M}$. Here $\sqrt{\alpha_{1}}=0.05$.}\label{fig3}
\end{minipage}
\begin{minipage}[t]{0.49\textwidth}
\centering
\includegraphics[width=1\linewidth]{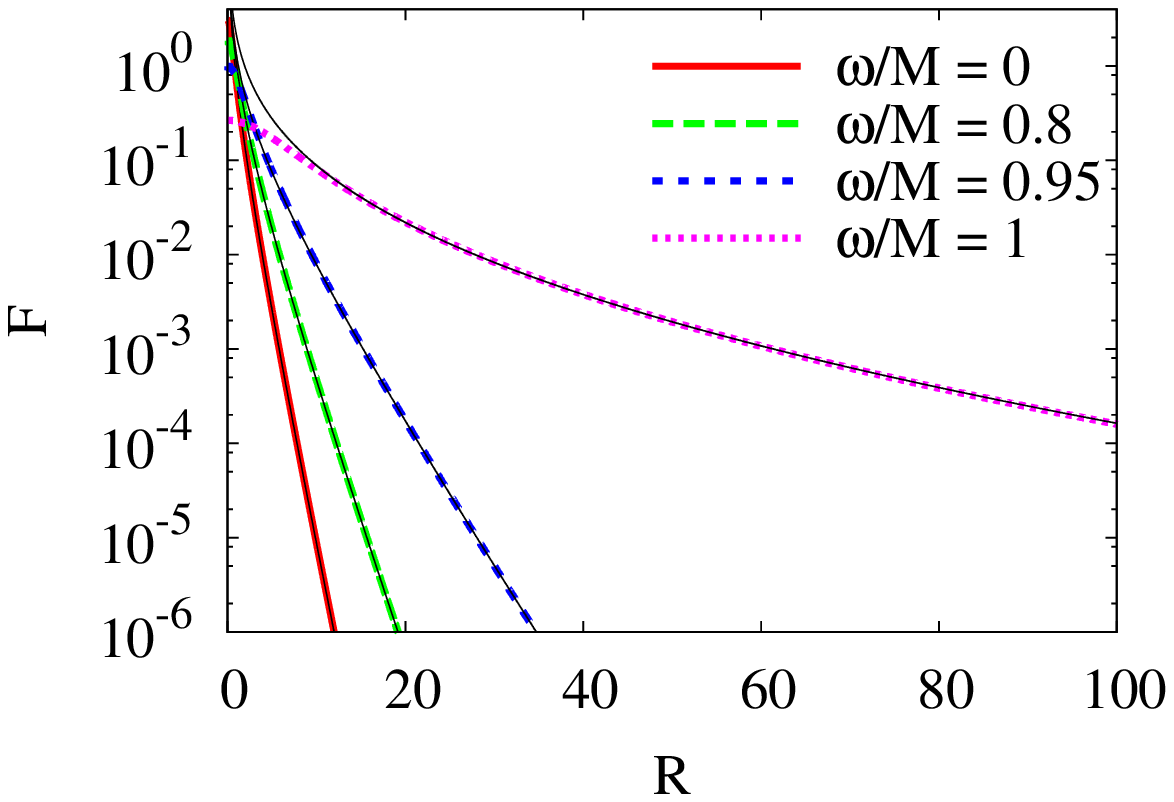}
\caption{Comparison of the scalar field profiles (thick lines) with the asymptotes defined by Eqs.~(\ref{frinfty1}),~(\ref{frinfty1a}) (thin lines) for different values of $\frac{\omega}{M}$. Here $\sqrt{\alpha_{1}}=0.05$.}\label{fig4}
\end{minipage}
\end{figure}
It is confirmed by the plots in Fig.~\ref{fig1}, demonstrating the existence of Q-balls with finite charges for $\frac{\omega}{M}=1$ for different values of the parameter $\alpha_{1}$.\footnote{The numerical analysis was also performed for the values of the parameter $\alpha_{1}$ larger than those used in Fig.~\ref{fig1}. However, no essential changes in the behavior of the $Q(\omega)$ dependencies were found.} We also performed the numerical analysis for $\frac{\omega}{M}>1$. No solutions of form (\ref{ans1})--(\ref{ans3}) with finite charge and energy were found.

In Fig.~\ref{fig2} one can find the $\frac{E}{Q}(Q)$ dependencies for different values of the parameter $\alpha_{1}$ in comparison with the nongauged case. Profiles of the scalar field for different values of $\frac{\omega}{M}$ and for $\sqrt{\alpha_{1}}=0.05$ are presented in Fig.~\ref{fig3}. We have also performed a comparison of the numerical solutions for the scalar field at large $R$ with the asymptotes defined by Eqs.~(\ref{frinfty1}),~(\ref{frinfty1a}). The result is presented in Fig.~\ref{fig4}, demonstrating a remarkable agreement with theoretical predictions.

\subsection{Model with the piecewise parabolic scalar field potential}
Now we consider the piecewise potential of form
\begin{equation}
V(f)=M^2f^2\,\theta\left(1-\frac{f^2}{v^2}\right)+M^2v^2\theta\left(\frac{f^2}{v^2}-1\right),
\label{potential}
\end{equation}
where $\theta$ is the Heaviside step function with the convention $\theta(0)=\frac{1}{2}$ (such piecewise potentials for the case of nongauged Q-balls were introduced in \cite{Rosen0} and thoroughly examined in \cite{Theodorakis:2000bz,Gulamov:2013ema}). Again, it is convenient to pass to the new variables
\begin{equation}
R=Mr,\qquad G(R)=\frac{1}{M}\,a(r),\qquad F(R)=\frac{1}{v}f(r).
\end{equation}
In these notations, the system of equations (\ref{eqg1a}), (\ref{eqg2a}) takes the form
\begin{eqnarray}\label{eqg1pcw}
2\alpha_{2}GF^2=\frac{1}{R}(RG)'',\\
\label{eqg2pcw} G^2F+\frac{1}{R}(RF)''-F\theta\left(1-F^2\right)=0,
\end{eqnarray}
where $\alpha_{2}=\frac{e^2v^2}{M^2}$, which is the only effective parameter in this system of equations. Again, since $\frac{1}{2f}\frac{dV}{df}\bigl|_{f=0}=M^{2}$, we will be looking for solutions such that $G(\infty)\le 1$, which corresponds to $\omega\le M$.

The charge of the Q-ball takes the form
\begin{equation}\label{chargepcw}
Q=\frac{v^2}{M^2}\,8\pi\int\limits_{0}^{\infty} GF^2R^2dR=\frac{v^2}{M^2}\,\tilde Q,
\end{equation}
whereas the energy is
\begin{eqnarray}\label{energypcw}
E=\frac{v^2}{M}\,4\pi\int\limits_{0}^{\infty}
\left(G^2F^2+\partial_{R}F\partial_{R}F+F^2\,\theta\left(1-F^2\right)+\theta\left(F^2-1\right)+\frac{1}{2\alpha_{2}}\partial_{R}G\partial_{R}G\right)R^2dR\\ \nonumber=\frac{v^2}{M}\,\tilde E.
\end{eqnarray}

\begin{figure}[!h]
\centering
\includegraphics[width=1\linewidth]{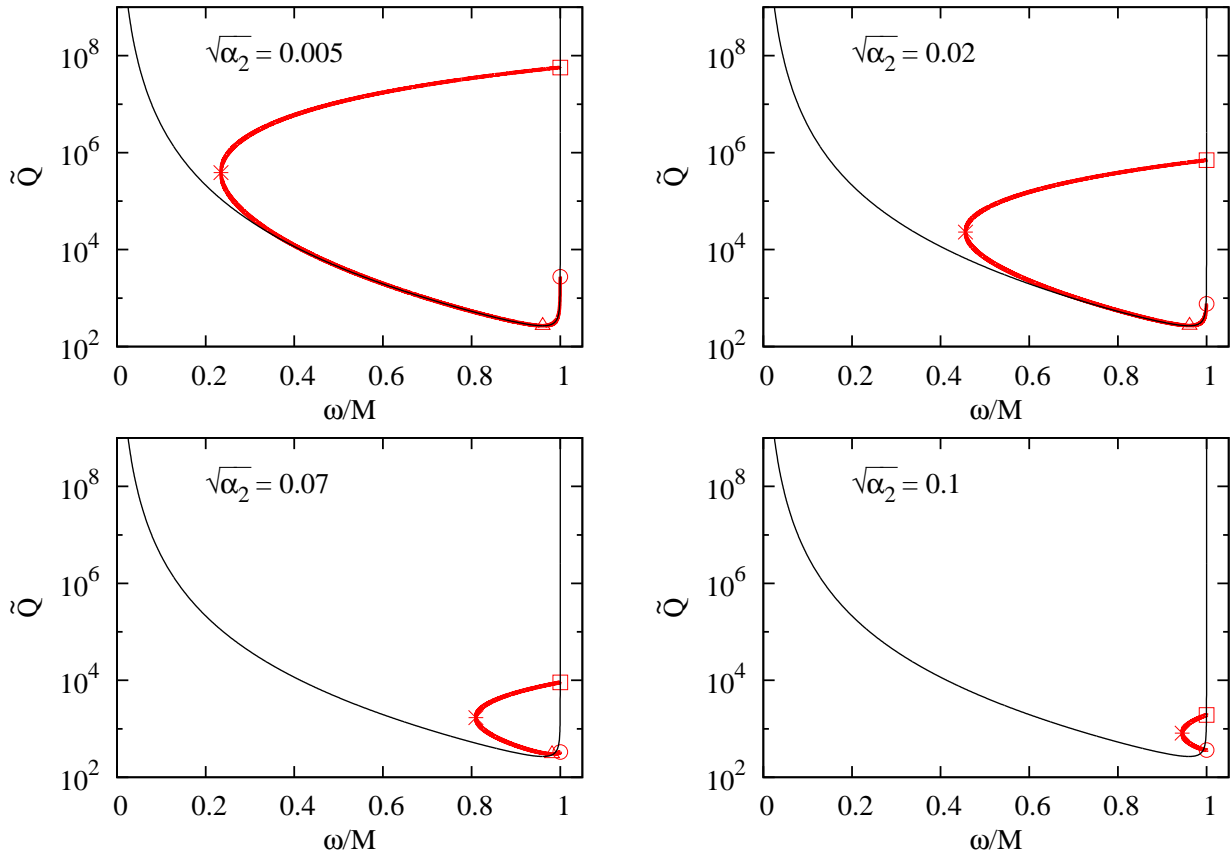}
\caption{$Q(\omega)$ for different values of the parameter $\alpha_{2}$ (thick lines). The thin lines stand for the nongauged case. The circles on the plots mark the points with $\frac{\omega}{M}=1$, the triangles mark the points with $\frac{dQ}{d\omega}=0$, the asterisks mark the points with $\frac{dQ}{d\omega}=\infty$, the boxes also mark the points with $\frac{\omega}{M}=1$.}\label{fig1b}
\end{figure}

In Fig.~\ref{fig1b} one can see several examples of $Q(\omega)$ diagrams for this model.
We see a completely unexpected behavior of the corresponding $Q(\omega)$ dependencies. First, we see that, contrary to the case of the previous model, now the parameter $\omega$ does not uniquely define the charge of the Q-ball. Indeed, except the Q-ball corresponding to a minimal value of $\omega$ (these points are marked by asterisks), there exist two Q-balls with different charges for each value of $\omega$. We have the following explanation of this fact.

Indeed, in the nongauged case the charge tends to infinity in the limits $\omega\to 0$ and $\omega\to M$ \cite{Gulamov:2013ema}. According to the results of Section~3, the nongauged limit $\omega\to M$ transforms into the Q-ball with $\omega=M$. This Q-ball corresponds to the lower points with $\frac{\omega}{M}=1$ in Fig.~\ref{fig1b} (these points are marked by the circles), and its existence is not surprising. Now let us consider another limit, namely $\omega\to 0$. From \cite{Rosen} we know that if $\omega=0$ for a gauged Q-ball, then $A_{0}\equiv 0$. Since there is not a Q-ball solution with the zero charge for $\omega=0$ in the nongauged case \cite{Gulamov:2013ema}, there should be no such solution in the gauged case too. It is also improbable that the charge of a gauged Q-ball tends to infinity in the limit $\omega\to 0$ --- the value of $\omega$ tends to zero, whereas $A_{0}(r)<0$ is a monotonically growing function such that $\omega+eA_{0}(r)>0$ for any $r$. Thus, $|eA_{0}(r)|<\omega\to 0$, whereas a solution for $A_{0}(r)$ should support the existence of a large charge. The latter situation seems to be unrealizable. So, the $Q(\omega)$ curve modifies with respect to the nongauged case in order to overcome this problem and to maintain its continuity, which is realized in Fig.~\ref{fig1b} --- the curve just turns back at the point of the minimal possible value of~$\omega$. At this point, the value of the charge behaves as $\tilde Q_{*}\sim\frac{1}{\alpha_{2}}\sim\frac{1}{e^{2}}$ for small values of $\alpha_{2}$.

\begin{figure}[!ht]
\centering
\includegraphics[width=1\linewidth]{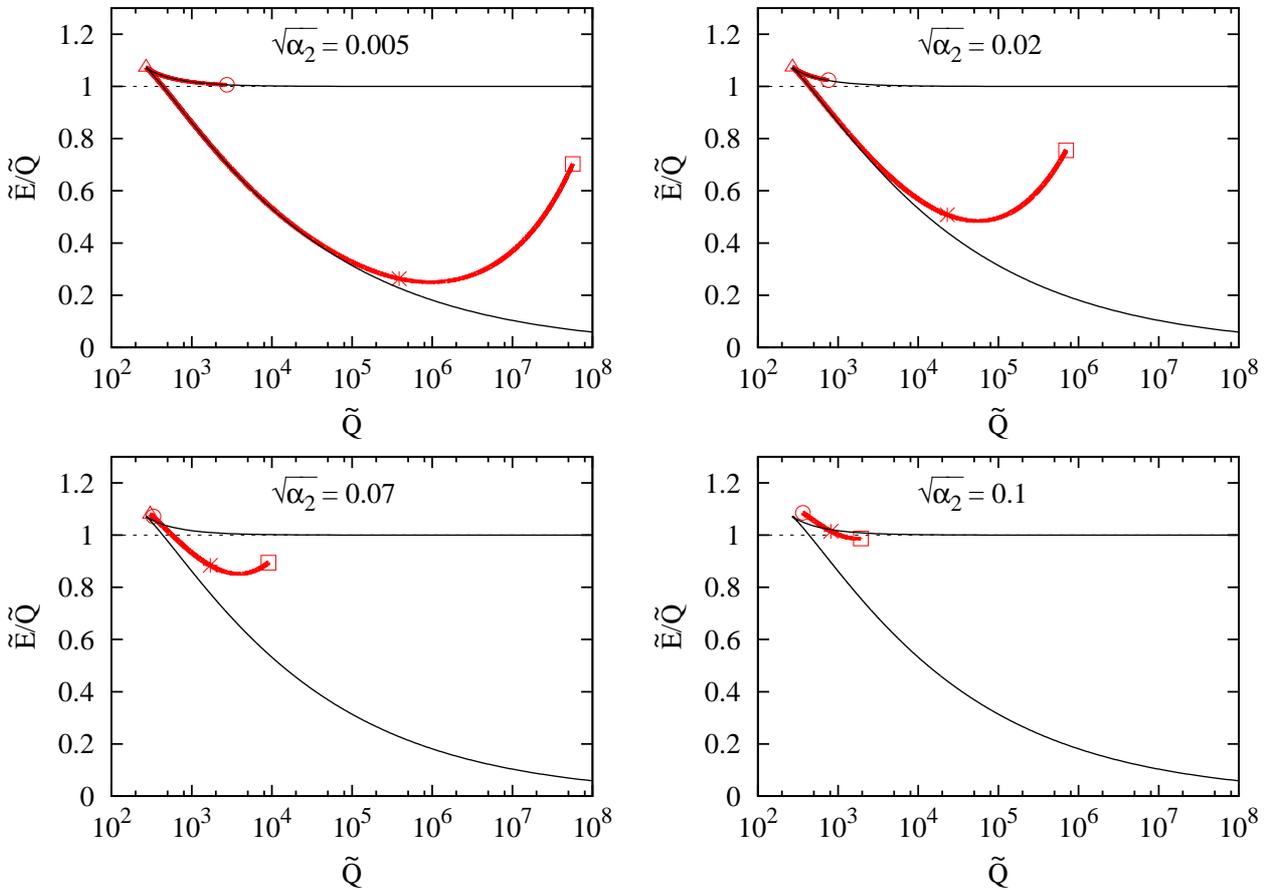}
\caption{$\frac{E}{Q}(Q)$ for different values of the parameter $\alpha_{2}$ (thick lines). The thin lines stand for the nongauged case. The circles on the plots mark the points with $\frac{\omega}{M}=1$, the triangles mark the points with $\frac{dQ}{d\omega}=0$, the asterisks mark the points with $\frac{dQ}{d\omega}=\infty$, the boxes also mark the points with $\frac{\omega}{M}=1$.}\label{fig2b}
\end{figure}

It is interesting to note that now it is the parameter $G(0)$ that uniquely characterizes a gauged Q-ball, not $\omega$ like in the nongauged case or even in the gauged case discussed in the previous subsection. We can also see that the curves in Fig.~\ref{fig1b} become smaller while increasing the value of the parameter $\alpha_{2}$. For $\alpha_{2}\gtrsim 0.013$ no gauged Q-balls were found. Analogous observation of nonexistence of gauged Q-ball for the values of the coupling constant larger than some critical value was made in \cite{Brihaye:2014gua}, where the model with supersymmetry motivated scalar field potential, which has the form similar to (\ref{potential}), was examined.

A remark is in order here. The results presented in this and in the previous subsections suggest that there exists a maximal possible charge of gauged Q-ball (of course, its value should depend on the model at hand). We think that it is not so in the general case; see, for example, \cite{Tamaki:2014oha}. However, this statement seems to be valid for models with $\frac{1}{2f}\frac{dV}{df}\bigl|_{f\to 0}\ne\infty$. It should be noted that this restriction is not connected with the restriction on the charge of {\em stable} gauged Q-balls \cite{Lee:1988ag}, which was shown to be incorrect in \cite{Gulamov:2013cra}.

\begin{figure}[!h]
\centering
\includegraphics[width=1\linewidth]{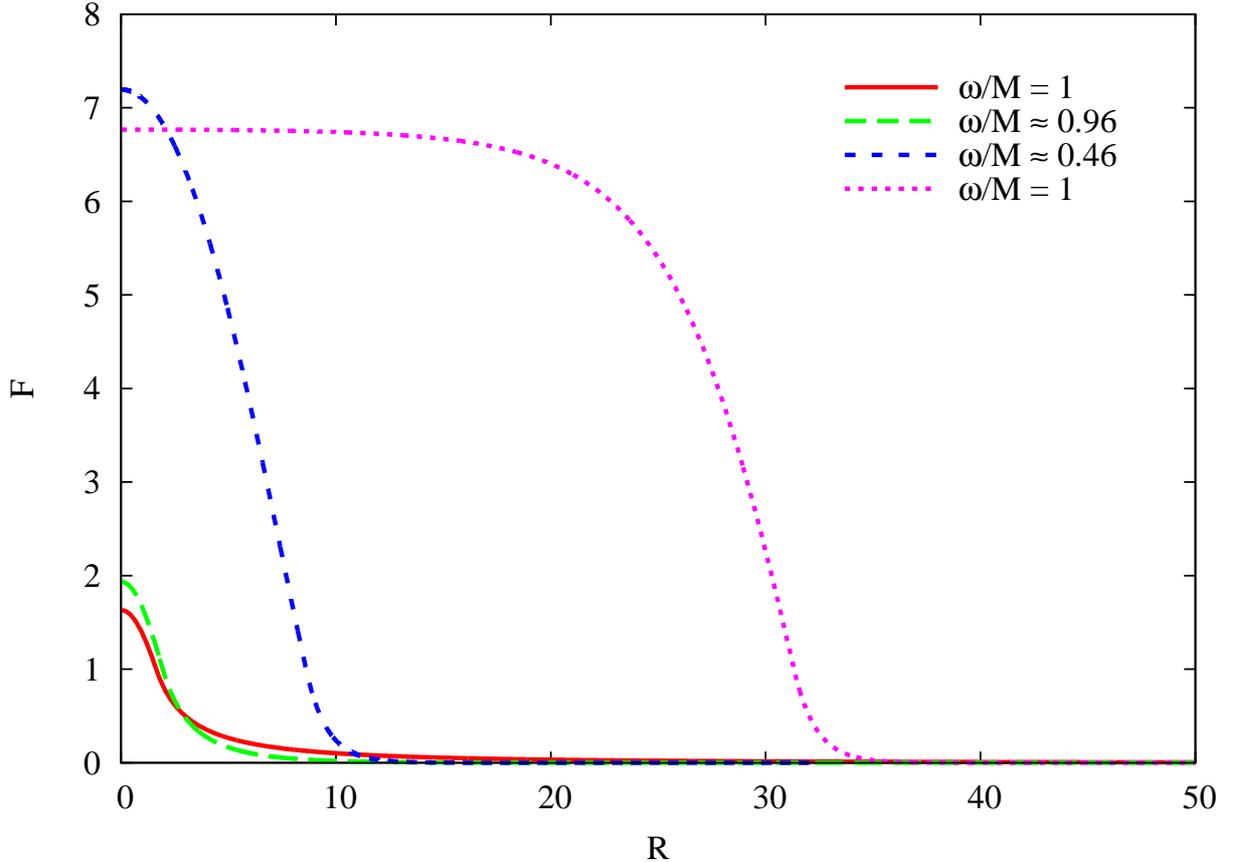}
\caption{Profiles of the scalar field for different values of $\frac{\omega}{M}$. Here $\sqrt{\alpha_{2}}=0.02$.}\label{fig3b}
\end{figure}
\begin{figure}[!ht]
\centering
\begin{minipage}[t]{0.49\textwidth}
\centering
\includegraphics[width=1\linewidth]{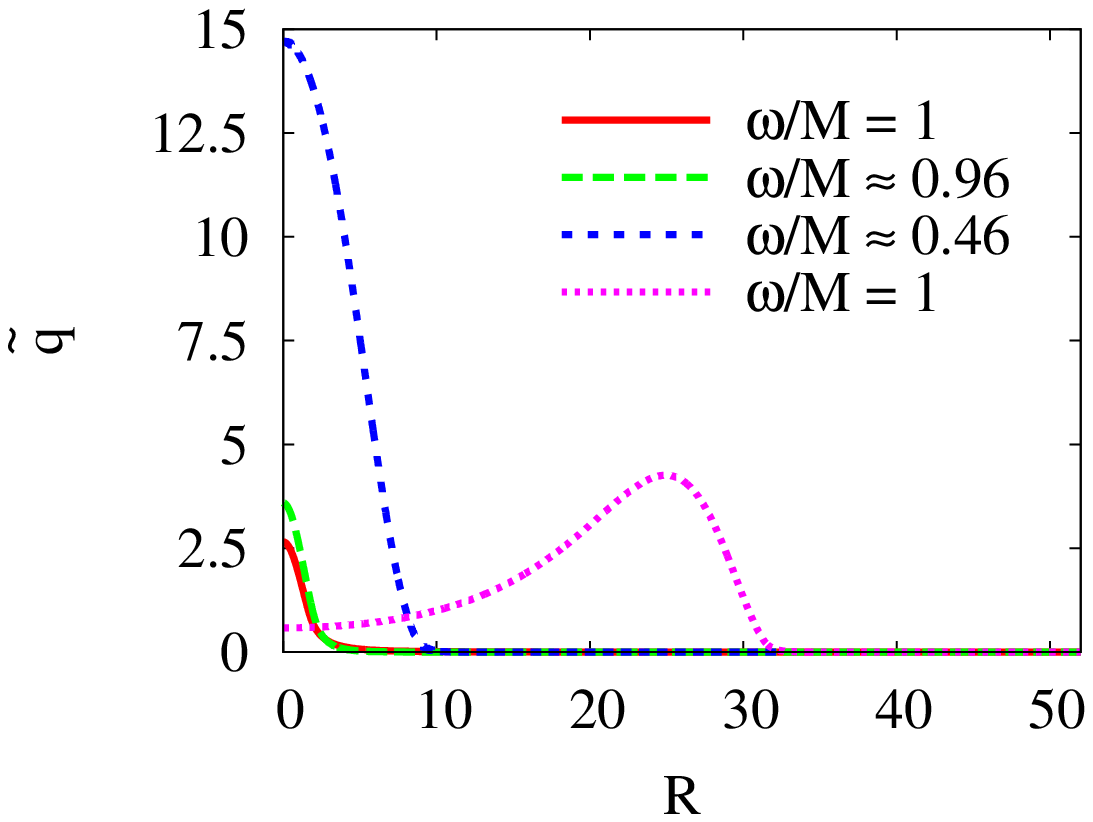}
\caption{Profiles of the effective charge density $\tilde q(R)$, $\sqrt{\alpha_{2}}=0.02$.}\label{fig4b}
\end{minipage}
\begin{minipage}[t]{0.49\textwidth}
\centering
\includegraphics[width=1\linewidth]{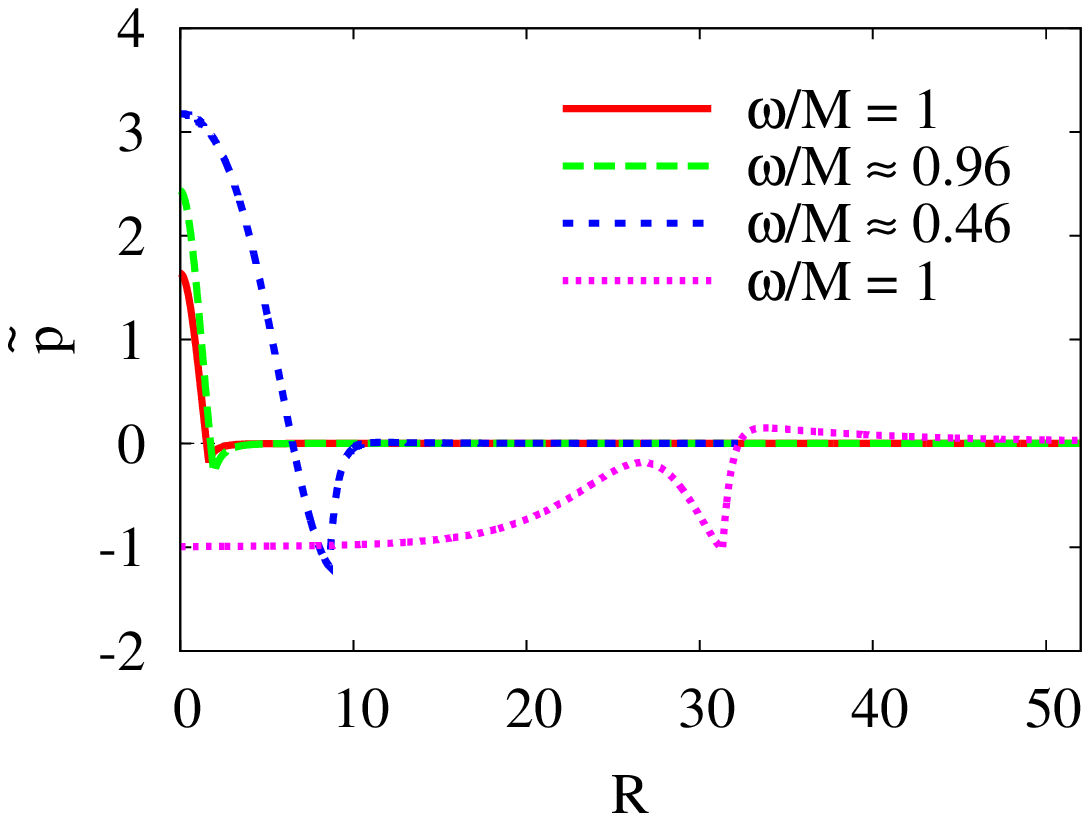}
\caption{Profiles of the effective pressure $\tilde p(R)$, $\sqrt{\alpha_{2}}=0.02$.}\label{fig5b}
\end{minipage}
\end{figure}

In Fig.~\ref{fig2b} one sees the $\frac{E}{Q}(Q)$ dependencies for different values of the parameter $\alpha_{2}$ in comparison with the nongauged case. The cusps on the corresponding curves, which are marked by triangles, correspond to the points with $\frac{dQ}{d\omega}=0$, which are also marked by triangles in Fig.~\ref{fig1b}. The existence of the cusps follows from the fact that $\frac{dE}{dQ}=\omega$ for gauged Q-balls \cite{Gulamov:2013cra}.

For the completeness, we also present the scalar field profiles (Fig.~\ref{fig3b}), the effective charge density $\tilde q(R)=GF^2$ (Fig.~\ref{fig4b}) and the effective pressure $\tilde p(R)$ (Fig.~\ref{fig5b}) for different values of $\frac{\omega}{M}$ and for $\sqrt{\alpha_{2}}=0.02$. The effective (dimensionless) pressure is defined in the standard way through the energy-momentum tensor \cite{Mai:2012yc} as
\begin{equation}
\tilde p=G^2F^2-\frac{1}{3}\left(\frac{dF}{dR}\right)^2+\frac{1}{6\alpha_{2}}\left(\frac{dG}{dR}\right)^2-\left(F^2\,\theta\left(1-F^2\right)+\theta\left(F^2-1\right)\right),
\end{equation}
one can show that the equality $\int_{0}^{\infty}\tilde p(R)R^{2}dR=0$ (the so-called ``von Laue condition'' \cite{Mai:2012yc}) fulfills in the general case (this equality was also used for the additional cross-check of our numerical results). The values of $\frac{\omega}{M}$ for these plots are chosen such that they approximately correspond to the marked points in Fig.~\ref{fig1b} in the clockwise order: $\frac{\omega}{M}=1$ --- circle, $\frac{\omega}{M}\approx 0.96$ --- triangle, $\frac{\omega}{M}\approx 0.46$ --- asterisk (the point of the turnover), $\frac{\omega}{M}=1$ --- box.

There exists a dip in the vicinity of $R=0$ on the charge density curve in Fig.~\ref{fig4b} for $\frac{\omega}{M}=1$, clearly indicating the repulsive nature of the electrostatic interaction in the gauged Q-ball. Such dips on the charge density curves appear for Q-balls from the upper part of the $Q(\omega)$ diagram (Fig.~\ref{fig1b}) for $\frac{\omega}{M}\gtrsim 0.506$. It is also interesting to note that, according to Fig.~\ref{fig5b}, the pressure can be negative even in the center of gauged Q-balls.

\section{Conclusion}
As it was demonstrated above, gauged Q-balls posses surprising properties which differ considerably from those of Q-balls in the nongauged case. Namely, it was shown that there may exist gauged Q-ball solutions with a finite charge even for $\omega=M$ (if $\frac{1}{2f}\frac{dV}{df}\bigl|_{f=0}=M^{2}\ne 0$), which is usually impossible in the nongauged case. The corresponding analytical considerations were supported by the numerical calculations in two models with different scalar field potentials. Moreover, numerical analysis shows that even when $Q\to\infty$ as $\omega\to 0$ in the nongauged case, the charge in the gauged case remains finite for all allowed values of $\omega$. As a consequence, the regions of allowed frequencies appear to be different from those in the nongauged case.

\section*{Acknowledgements}
The authors are grateful to D.G.~Levkov, M.V.~Libanov and S.V.~Troitsky for valuable discussions. The work was supported by grant 14-02-31384 of the Russian Foundation for Basic Research (analytical analysis of Section~3) and by grant 14-22-00161 of the Russian Science Foundation (numerical analysis of Section~4).

\section*{Appendix A: The behavior of $f(r)$ for $\sqrt{M^2-\omega^2}\,r\gg 1$}
Let us take the integral representation of the confluent hypergeometric function of the second kind $U(b,c,z)$, which has the form \cite{AS}
\begin{equation}
U(b,c,z)=\frac{1}{\Gamma(b)}\int\limits_{0}^{\infty}\textrm{e}^{-zt}t^{b-1}(1+t)^{c-b-1}dt.
\end{equation}
For solution (\ref{frinfty}) we get
\begin{equation}
f(r)\sim \textrm{e}^{-\sqrt{M^2-\omega^2}\,r}\int\limits_{0}^{\infty}\textrm{e}^{-2\sqrt{M^2-\omega^2}\,rt}t^{\frac{\omega e^2Q}{4\pi\sqrt{M^2-\omega^2}}}(1+t)^{-\frac{\omega e^2Q}{4\pi\sqrt{M^2-\omega^2}}}dt.
\end{equation}
Let us change the variable $t$ such that $\sqrt{M^2-\omega^2}\,rt=\tilde t$. We get
\begin{equation}\label{appaint}
f(r)\sim \frac{\textrm{e}^{-\sqrt{M^2-\omega^2}\,r}}{r}\int\limits_{0}^{\infty}\textrm{e}^{-2\tilde t}\left(\frac{\tilde t}{\sqrt{M^2-\omega^2}\,r+\tilde t}\right)^{\frac{\omega e^2Q}{4\pi\sqrt{M^2-\omega^2}}}d\tilde t.
\end{equation}
It is clear that, due to the exponential suppression, the main contribution to the integral in (\ref{appaint}) is achieved in the region $\tilde t\sim 1$. Thus, for $\sqrt{M^2-\omega^2}\,r\gg 1$, we can rewrite the integral in (\ref{appaint}) as
\begin{eqnarray}\label{appaint1}
\int\limits_{0}^{\infty}\textrm{e}^{-2\tilde t}\left(\frac{\tilde t}{\sqrt{M^2-\omega^2}\,r+\tilde t}\right)^{\frac{\omega e^2Q}{4\pi\sqrt{M^2-\omega^2}}}d\tilde t\approx
\int\limits_{0}^{\infty}\textrm{e}^{-2\tilde t}\left(\frac{\tilde t}{\sqrt{M^2-\omega^2}\,r}\right)^{\frac{\omega e^2Q}{4\pi\sqrt{M^2-\omega^2}}}d\tilde t\\ \nonumber=\left(\frac{1}{\sqrt{M^2-\omega^2}\,r}\right)^{\frac{\omega e^2Q}{4\pi\sqrt{M^2-\omega^2}}}\int\limits_{0}^{\infty}\textrm{e}^{-2\tilde t}{\tilde t}^{\frac{\omega e^2Q}{4\pi\sqrt{M^2-\omega^2}}}d\tilde t.
\end{eqnarray}
Finally, for (\ref{appaint}) we obtain (up to a constant depending on $\omega$)
\begin{equation}
f(r)\sim \frac{\textrm{e}^{-\sqrt{M^2-\omega^2}\,r}}{r^{1+\frac{\omega e^2Q}{4\pi\sqrt{M^2-\omega^2}}}},
\end{equation}
which coincides with (\ref{frinfty1}).

\section*{Appendix B: Correspondence between the scalar field solutions for $\omega<M$ and $\omega=M$}
Let us again take the integral representation of the confluent hypergeometric function of the second kind $U(b,c,z)$. We will be interested in the limit $\omega\to M$ for a fixed $r$. In this case (\ref{appaint}) can be rewritten as
\begin{equation}\label{appaint2}
f(r)\sim \frac{1}{r}\lim\limits_{\omega\to M}\int\limits_{0}^{\infty}\textrm{e}^{-2\tilde t}\left(\frac{\tilde t}{\sqrt{M^2-\omega^2}\,r+\tilde t}\right)^{\frac{\omega e^2Q}{4\pi\sqrt{M^2-\omega^2}}}d\tilde t.
\end{equation}
In the limit $\omega\to M$ the term $\frac{\tilde t}{\sqrt{M^2-\omega^2}\,r+\tilde t}$ can be represented as
\begin{equation}
\frac{\tilde t}{\sqrt{M^2-\omega^2}\,r+\tilde t}=1-\frac{\sqrt{M^2-\omega^2}\,r}{\sqrt{M^2-\omega^2}\,r+\tilde t}\approx 1-\frac{\sqrt{M^2-\omega^2}\,r}{\tilde t}
\end{equation}
Thus, we get
\begin{eqnarray}\label{appaint3}
\lim\limits_{\omega\to M}\int\limits_{0}^{\infty}\textrm{e}^{-2\tilde t}\left(\frac{\tilde t}{\sqrt{M^2-\omega^2}\,r+\tilde t}\right)^{\frac{\omega e^2Q}{4\pi\sqrt{M^2-\omega^2}}}d\tilde t\approx\lim\limits_{\omega\to M}\int\limits_{0}^{\infty}\textrm{e}^{-2\tilde t}\left(1-\frac{\sqrt{M^2-\omega^2}\,r}{\tilde t}\right)^{\frac{\omega e^2Q}{4\pi\sqrt{M^2-\omega^2}}}d\tilde t\\ \nonumber=\lim\limits_{\omega\to M}\int\limits_{0}^{\infty}\textrm{e}^{-2\tilde t}\left(\left(1-\frac{\sqrt{M^2-\omega^2}\,r}{\tilde t}\right)^{-\frac{\tilde t}{\sqrt{M^2-\omega^2}\,r}}\right)^{-\frac{\omega e^2Q\,r}{4\pi\tilde t}}d\tilde t=
\int\limits_{0}^{\infty}\textrm{e}^{-2\tilde t}\textrm{e}^{-\frac{Me^2Q\,r}{4\pi\tilde t}}d\tilde t.
\end{eqnarray}
Now let us define $y=\sqrt{\frac{2Me^2Q}{\pi}\,r}$, $x=\sqrt{\frac{8\pi}{Me^2Q\,r}}\,\tilde t$. In these notations, the last integral in (\ref{appaint3}) can be rewritten as
\begin{eqnarray}\label{appaint4}
\int\limits_{0}^{\infty}\textrm{e}^{-2\tilde t}\textrm{e}^{-\frac{Me^2Q\,r}{4\pi\tilde t}}d\tilde t=\sqrt{\frac{M e^2Q\,r}{8\pi}}\int\limits_{0}^{\infty}\textrm{e}^{-\frac{y}{2}\left(x+\frac{1}{x}\right)}dx\\ \nonumber=\sqrt{\frac{M e^2Q\,r}{8\pi}}\left(\int\limits_{0}^{1}\textrm{e}^{-\frac{y}{2}\left(x+\frac{1}{x}\right)}dx+\int\limits_{1}^{\infty}\textrm{e}^{-\frac{y}{2}\left(x+\frac{1}{x}\right)}dx\right).
\end{eqnarray}
Now let us make the redefinition $x\to\frac{1}{x}$ in the last integral in (\ref{appaint4}). We get
\begin{eqnarray}\label{appaint5}
\sqrt{\frac{M e^2Q\,r}{8\pi}}\left(\int\limits_{0}^{1}\textrm{e}^{-\frac{y}{2}\left(x+\frac{1}{x}\right)}dx+
\int\limits_{1}^{\infty}\textrm{e}^{-\frac{y}{2}\left(x+\frac{1}{x}\right)}dx\right)=
\sqrt{\frac{M e^2Q\,r}{8\pi}}\int\limits_{0}^{1}\textrm{e}^{-\frac{y}{2}\left(x+\frac{1}{x}\right)}\left(\frac{1+x^{2}}{x^{2}}\right)dx.
\end{eqnarray}
It is convenient to introduce the new variable $2w=x+\frac{1}{x}$, which leads, according to the limits of integration, to $x=w-\sqrt{w^{2}-1}$. The integral in (\ref{appaint5}) can be rewritten as
\begin{eqnarray}\label{appaint6}
\sqrt{\frac{M e^2Q\,r}{8\pi}}\int\limits_{0}^{1}\textrm{e}^{-\frac{y}{2}\left(x+\frac{1}{x}\right)}\left(\frac{1+x^{2}}{x^{2}}\right)dx=\sqrt{\frac{M e^2Q\,r}{8\pi}}\int\limits_{1}^{\infty}\textrm{e}^{-wy}\left(\frac{2w}{\sqrt{w^2-1}}\right)dw\\ \nonumber=\sqrt{\frac{M e^2Q\,r}{2\pi}}\,y\int\limits_{1}^{\infty}\textrm{e}^{-wy}\sqrt{w^2-1}\,dw,
\end{eqnarray}
where we have performed integration by parts in the last step. Recalling the integral representation of the modified Bessel function of the second kind $K_{1}(y)$, which has the form \cite{AS}
\begin{eqnarray}\label{intBessel}
K_{1}(y)=\frac{\sqrt{\pi}}{2\Gamma\left(\frac{3}{2}\right)}\,y\int\limits_{1}^{\infty}\textrm{e}^{-wy}\sqrt{w^2-1}\,dw,
\end{eqnarray}
combining formulas (\ref{appaint2}), (\ref{appaint6}), (\ref{intBessel}) and taking into account the definition of $y$, we arrive at
\begin{equation}
f(r)\sim \frac{K_{1}\left(\sqrt{\frac{2Me^2Q}{\pi}\,r}\right)}{\sqrt{r}},
\end{equation}
which obviously corresponds to (\ref{frinfty2}).


\begin{thebibliography}{99}
\bibitem{Rosen0}
G. ~Rosen, J.\ Math.\ Phys. {\bf 9} (1968) 996.

\bibitem{Coleman:1985ki}
S.~R.~Coleman, Nucl.\ Phys.\ B {\bf 262} (1985) 263 [Erratum-ibid.\ B {\bf 269} (1986) 744].

\bibitem{Rosen}
G.~Rosen, J. Math. Phys. {\bf 9} (1968) 999.

\bibitem{Lee:1988ag}
K.~-M.~Lee, J.~A.~Stein-Schabes, R.~Watkins and L.~M.~Widrow, Phys.\ Rev.\ D {\bf 39} (1989) 1665.

\bibitem{BF}
V.~Benci and D.~Fortunato, J. Math. Phys. {\bf 52} (2011) 093701.

\bibitem{BF1}
V.~Benci and D.~Fortunato, Chaos Solitons Fractals {\bf 58} (2014) 1.

\bibitem{Gulamov:2013cra}
I.~E.~Gulamov, E.~Y.~Nugaev and M.~N.~Smolyakov, Phys.\ Rev.\ D {\bf 89} (2014) 085006.

\bibitem{Lee:1991bn}
C.~H.~Lee and S.~U.~Yoon, Mod.\ Phys.\ Lett.\ A {\bf 06} (1991) 1479.

\bibitem{Arodz:2008nm}
H.~Arodz and J.~Lis, Phys.\ Rev.\ D {\bf 79} (2009) 045002.

\bibitem{Dzhunushaliev:2012zb}
V.~Dzhunushaliev and K.~G.~Zloshchastiev, Central Eur.\ J.\ Phys.\ {\bf 11} (2013) 325.

\bibitem{Tamaki:2014oha}
T.~Tamaki and N.~Sakai, Phys.\ Rev.\ D {\bf 90} (2014) 085022.

\bibitem{Brihaye:2014gua}
Y.~Brihaye, V.~Diemer and B.~Hartmann, Phys.\ Rev.\ D {\bf 89} (2014) 084048.

\bibitem{Multamaki:1999an}
T.~Multamaki and I.~Vilja, Nucl.\ Phys.\ B {\bf 574} (2000) 130.

\bibitem{Derrick}
G.~H.~Derrick, J. Math. Phys. {\bf 5} (1964) 1252.

\bibitem{Smolyakov}
M.~N.~Smolyakov, J. Phys. A: Math. Theor. {\bf 43} (2010) 455202.

\bibitem{AD}
D.L.T.~Anderson, G.H.~Derrick, J. Math. Phys. {\bf 11} (1970) 1336.

\bibitem{Theodorakis:2000bz}
S.~Theodorakis, Phys.\ Rev.\ D {\bf 61} (2000) 047701.

\bibitem{Gulamov:2013ema}
I.~E.~Gulamov, E.~Y.~Nugaev and M.~N.~Smolyakov, Phys.\ Rev.\ D {\bf 87} (2013) 085043.

\bibitem{Mai:2012yc}
M.~Mai and P.~Schweitzer, Phys.\ Rev.\ D {\bf 86} (2012) 076001.

\bibitem{AS}
M.~Abramowitz, I.~A.~Stegun, ``Handbook of Mathematical Functions with Formulas, Graphs, and Mathematical Tables'', Dover Publications, New York, 1972.

\end{thebibliography}
\end{document}